\documentclass{appolb}
\usepackage{epsfig}

\usepackage{bm}%
\usepackage[colorlinks=true,linkcolor=red]{hyperref}
\usepackage{amsmath}
\usepackage{amsfonts}
\usepackage{amssymb}

\usepackage{graphicx}

\begin{document}
\title{Transmission Eigenvalue Densities and Moments in Chaotic Cavities from Random Matrix Theory%
}
\author{Pierpaolo Vivo
\address{School of Information Systems, Computing \& Mathematics\\Brunel
University\\ Uxbridge, Middlesex, UB8 3PH\\(United Kingdom)} \and
Edoardo Vivo
\address{Universit\`{a} degli Studi di Parma\\
Dipartimento di Fisica Teorica\\Viale G.P. Usberti n.7/A (Parco
Area delle Scienze), Parma\\ (Italy)}} \maketitle
\begin{abstract}
We point out that the transmission eigenvalue density and higher
order correlation functions in chaotic cavities for an arbitrary
number of incoming and outgoing leads $(N_1,N_2)$ are analytically
known from the Jacobi ensemble of Random Matrix Theory. Using this
result and a simple linear statistic, we give an exact and
non-perturbative expression for moments of the form $\langle
\lambda_1^m\rangle$ for $m>-|N_1-N_2|-1$ and $\beta=2$, thus
improving the existing results in the literature. Secondly, we
offer an independent derivation of the average density and higher
order correlation functions for $\beta=2,4$ which does not make
use of the orthogonal polynomials technique. This result may be
relevant for an efficient numerical implementation avoiding
determinants.
\end{abstract}

\section{Introduction}
Conductance in mesoscopic systems is currently a very active area
of research, both from the theoretical and the experimental point
of view. In the scattering theory framework
\cite{ya}\cite{beenakker}, for the case of a chaotic cavity with
$N_1$ and $N_2$ channels in each of the two attached leads, the
fluctuations of the transmission eigenvalues of the conductor are
effectively provided by a random matrix with appropriate
symmetries \cite{baranger}\cite{jalabert}. More specifically, the
Dyson index $\beta$ of the ensemble acquires the values $1$ or $2$
according to the presence or absence of time-reversal symmetry, or
$4$ in the case of spin-flip symmetry.

Several quantities of interest for the experiments, such as the
conductance and the average shot noise, may be derived from the
knowledge of the transmission eigenvalues $\{\lambda_i\}$. Those
are defined as the singular values of a transmission matrix $t$,
which in turn is a $N_1\times N_2$ off-diagonal block of a
$\tilde{N}\times \tilde{N}$ unitary scattering matrix (where
$\tilde{N}=N_1+N_2$) \cite{buttiker}. In the case of chaotic
cavities considered below, $\{\lambda_j\}$ are correlated random
variables between $0$ and $1$.

Suppose that one is interested in computing the average shot noise
$\langle P\rangle$, where:
\begin{equation}\label{AverageShotNoise}
  P=P_0\sum_{p=1}^N \lambda_p (1-\lambda_p),\qquad
  N=\min(N_1,N_2),
\end{equation}
$P_0$ being a constant related to the physical properties of the
conductor \cite{buttiker}\cite{khlus}\cite{lesovik}. Until 2005,
results for $\langle P\rangle$ were known only in the limiting
cases $N_{1,2}\gg 1$
\cite{beenakker}\cite{nazarov}\cite{schanz}\cite{whitney},
$N_1=N_2=1$ \cite{pedersen} or few open channels \cite{araujo}.
Very recently, a compact form has been found for $\langle
P\rangle/P_0 $ using two different methods, based on a
semiclassical expansion \cite{braun} and on recurrence relations
for the Selberg integral \cite{savin} (see also \cite{bulgakov}
for an alternative derivation). The latter nicely exploits the
remarkably simple expression for the joint probability density
(jpd) of
transmission eigenvalues:
\begin{equation}\label{jpd}
  P_\beta(\lambda_1,\ldots,\lambda_N)=\mathcal{N}_\beta^{-1}\prod_{j<k}|\lambda_j-\lambda_k|^\beta
  \prod_{i=1}^N
  \lambda_i^{\frac{\beta}{2}(|N_2-N_1|+1)-1},\qquad
  0\leq\lambda_j\leq 1,
\end{equation}
where the normalization constant is given by
\cite{savin}\cite{mehta}:
\begin{equation}\label{NormConstant}
  \mathcal{N}_\beta=\prod_{j=0}^{N-1}\frac{\Gamma\left(1+\frac{\beta}{2}+j\frac{\beta}{2}\right)\Gamma\left(\frac{\beta}{2}(|N_2-N_1|+1)+j\frac{\beta}{2}\right)
  \Gamma\left(1+j\frac{\beta}{2}\right)}{\Gamma\left(1+\frac{\beta}{2}\right)\Gamma\left(\frac{\beta}{2}(|N_2-N_1|+1)+1+(N+j-1)\frac{\beta}{2}\right)}.
\end{equation}
A few comments about \eqref{jpd} are in order. The jpd in
\eqref{jpd} is stated in \cite{beenakker} without proof and
attributed to Brouwer. A formal proof has been given (using three
different methods) by Forrester \cite{forrcond} in 2006, where the
author also highlighted the connection with the jpd of the Jacobi
ensemble of random matrices \cite{mehta}\cite{forrester}. In fact,
one observes that the change of variables $y_j=1-2\lambda_j$
brings \eqref{jpd} to the form:
\begin{equation}\label{jpdJacobi}
  P_\beta(y_1,\ldots,y_N)=\mathcal{\tilde{N}}_\beta^{-1}\prod_{j<k}|y_j-y_k|^\beta
  \prod_{i=1}^N
  (1-y_i)^{\frac{\beta}{2}(|N_2-N_1|+1)-1},\qquad
  -1\leq y_j\leq 1,
\end{equation}
allowing to use the machinery and results already known from
Random Matrix Theory.

In particular, the average density of transmission eigenvalues:
\begin{equation}\label{AverageDensity}
  \rho_\beta(\lambda;N_1,N_2)=\langle\sum_{i=1}^N
  \delta(\lambda-\lambda_i)\rangle=N\int_{[0,1]^{N-1}}~d\lambda_2\ldots
  d\lambda_N P_\beta(\lambda,\lambda_2\ldots,\lambda_N)
\end{equation}
is of interest for computing linear statistics, i.e. observables
of the form $\langle \mathrm{tr} f(t t^\dagger)\rangle$:
\begin{equation}\label{linear statistics}
\langle \mathrm{tr} f(t t^\dagger)\rangle =\int_0^1 dx
\rho_\beta(x;N_1,N_2)f(x).
\end{equation}
The moments of the form $\langle \lambda_1^m\rangle$ for a real
number $m$ can be computed in principle from the knowledge of the
average density as:
\begin{equation}\label{Moments}
\langle \lambda_1^m\rangle=\int_0^1 dx x^m \rho_\beta(x;N_1,N_2),
\end{equation}
where the range for $m$ is constrained by the convergence of the
integral. The first two moments are directly related to the
normalized conductance $(G/G_0=\langle \lambda_1\rangle)$ thanks
to the Landauer-B\"{u}ttiker formula, and to the already mentioned
shot noise $(P/P_0=\langle \lambda_1\rangle-\langle
\lambda_1^2\rangle)$. A refined semiclassical treatment of the
former can be found in \cite{richter}.

Surprisingly, the connection with the Jacobi ensemble has not been
fully appreciated so far, with the consequence that the average
spectral density $\rho_\beta(\lambda;N_1,N_2)$ for \emph{finite}
and \emph{arbitrary} number of open channels $(N_1,N_2)$ is still
deemed unknown (see e.g. \cite{savin}\cite{savin2}). On the other
hand, the density is known in the above mentioned limiting cases
\cite{beenakker}\cite{baranger}\cite{jalabert}\cite{nazarov}\cite{araujo}\cite{brouwer}.

In the mesoscopic literature the Jacobi ensemble is mentioned in
the paper by Ara\'{u}jo and Mac\^{e}do \cite{araujo}, where the
authors derived the average density of transmission eigenvalues
\emph{for a small number of open channels and $\beta=2$} using an
auxiliary non-linear sigma model. Their result reads:
\begin{equation}\label{AraujoMacedo}
  \rho_2(\lambda;N_1,N_2<11)=\lambda^\mu \sum_{n=0}^{N-1}(2n+\mu+1)\{P_n^{(\mu,0)}(1-2\lambda)\}^2
\end{equation}
where $\mu=\frac{\beta}{2}(|N_2-N_1|+1)-1=|N_2-N_1|$,
$N=\min(N_1,N_2)$ and $P_n^{(\alpha,\beta)}(y)$ is a Jacobi
polynomial.

The authors state in \cite{araujo}:
\begin{quotation}
...we believe (although we have no formal proof) that Eq.
\eqref{AraujoMacedo} is valid for arbitrary $N_1$ and $N_2$. This
result is consistent with the random-matrix approach of Ref.
\cite{baranger}\cite{jalabert}, which predicts for the same system
a joint distribution of transmission eigenvalues given by the
Jacobi ensemble, from which Eq. \eqref{AraujoMacedo} can be
derived. We have thus found independent evidence for the
application of the Jacobi ensemble in this problem.
\end{quotation}
However, the invoked references \cite{baranger}\cite{jalabert} do
not mention the Jacobi ensemble, and work out the only case
$N_1=N_2$. More precisely:
\begin{enumerate}
  \item Ref. \cite{baranger} reports the jpd \eqref{jpd} \emph{restricted to the case $N_1=N_2$ and
  $\beta=1,2$}. For the case $\beta=2$, the term $\lambda_i^{\frac{\beta}{2}(|N_2-N_1|+1)-1}$ in the jpd \eqref{jpd}
  then disappears, making the use of Legendre polynomials appropriate. For this subcase, the authors derive the average
  density and the 2-point function, and finally take the large $N_1=N_2$ limit to get
  the smoothed macroscopic density $\rho_2(\lambda;N_1=N_2\gg 1)\approx
  N/\pi\sqrt{\lambda(1-\lambda)}$;
  \item Ref. \cite{jalabert} deals with all symmetry classes
  $\beta=1,2,4$ and considers the two cases $N_1=N_2\gg 1$ or
  $N_1=N_2=1$. In the first subcase, the authors derive some quantities of interest with the use
  of a Coulomb gas approach after the change of variable $\lambda_i =
  1/(1+y_i)$, $y_i\in [0,\infty)$.
\end{enumerate}

We wish to clarify that the average density of transmission
eigenvalues for \textit{any} $N_1$ and $N_2$ is exactly given by
the density of the Jacobi ensemble, where the argument of the
Jacobi polynomials is $1-2\lambda$ (i.e. nothing but
\eqref{AraujoMacedo}, for $\beta=2$), and this result descends
from the application of the standard Orthogonal Polynomial
Technique \cite{mehta}\cite{nagao} to the (modified) jpd
\eqref{jpdJacobi}. In fact, the Jacobi polynomials $P_n^{(\mu,0)}$
appearing in \eqref{AraujoMacedo} are precisely \emph{the}
orthogonal polynomials over $[-1,1]$ with respect to the weight
$(1-y)^\mu$ in \eqref{jpdJacobi}. The cases $\beta=1$ and
$\beta=4$ are more complicated, but can be tackled in the same
framework (see \cite{ghosh} and references therein). Also, $n$-th
order correlation functions can be derived for all three symmetry
classes \cite{mehta}\cite{ghosh}. For example, for $\beta=2$ one
defines the kernel (see \cite{mehta}, Sections 5.7 and 19.1):
\begin{equation}\label{kernel}
  K_N(x,y)=x^{\mu/2}y^{\mu/2}\sum_{n=0}^{N-1}(2n+\mu+1)P_n^{(\mu,0)}(1-2x)P_n^{(\mu,0)}(1-2y),
\end{equation}
and the $n$-th order correlation function is written in terms of
the $(n\times n)$ determinant:
\begin{align}\label{corr}
  \nonumber\rho_2(\lambda_1,\ldots,\lambda_n) &=\frac{N!}{(N-n)!}\int_{[0,1]^{N-n}}d\lambda_{n+1}\ldots d\lambda_N
  P_\beta(\lambda_1,\ldots,\lambda_N)\\
  &=\det[K_N(\lambda_j,\lambda_k)]_{1\leq j,k\leq n}
\end{align}
In particular, the average spectral density (1-point function) is
exactly given by:
\begin{equation}\label{spectralexact}
  \rho_2(\lambda;N_1,N_2)=\lambda^\mu \sum_{n=0}^{N-1}(2n+\mu+1)\{P_n^{(\mu,0)}(1-2\lambda)\}^2
\end{equation}
extending the result \eqref{AraujoMacedo} to an arbitrary number
of open channels. In Appendix B, we will show that for $\mu\to 0$
eq. \eqref{spectralexact} recovers the result by Baranger and
Mello \cite{baranger}.

The purpose of this paper is thus twofold:
\begin{itemize}
  \item having clarified the role of the Jacobi ensemble, and the known results for its spectral density
  for arbitrary $N_1$ and $N_2$, we give a closed form expression for moments of the
  form $\langle \lambda_1^m\rangle$ for $m>-|N_2-N_1|-1$ and $\beta=2$ through a simple integration
  over the average density (linear statistic). The formula is exact
  and non-perturbative, and extends previous results in the
  literature (Section \ref{Sect2}).
  \item exploiting a less known result by Kaneko, we give an
  alternative representation for the average density and higher
  order correlation functions for $\beta=2,4$ in terms of hypergeometric functions
  of a matrix argument. Thanks to a recent algorithmic progress,
  this result may prove useful for a numerical implementation
  which avoids the use of determinants (or quaternion
  determinants) (Section \ref{sect3}).
\end{itemize}

\section{A closed form expression for moments
}\label{Sect2} For simplicity, we consider again the $\beta=2$
case as in \cite{araujo}. The moments $\langle\lambda_1^m\rangle$
can be computed as a simple linear statistic on the transmission
eigenvalues:
\begin{equation}\label{LinearStatistics}
  \langle\lambda_1^m\rangle=\int_0^1 dx x^m \rho_{2}(x;N_1,N_2)
\end{equation}
Known results about $\langle\lambda_1^m\rangle$ include:
\begin{enumerate}
  \item \emph{Approximate} evaluation \emph{for all positive integer} $m$ (but valid in the regime $N_1,N_2\gg
  1$) \cite{novaes}\cite{berkolaiko}; also, the generating function for all moments
  in this limit was first computed in \cite{brouwer}.
  \item \emph{Exact} evaluation (valid for all $N_1,N_2,\beta$) but only
  up to $m=4$ (see \cite{savin2} and references therein).
\end{enumerate}
Assuming $N_1<N_2$ without loss of generality, we can use
\eqref{spectralexact} and \eqref{LinearStatistics} with $N_1=N$
and $N_2=\mu+N$:
\begin{equation}\label{LinearStatistics}
  \langle\lambda_1^m\rangle=\sum_{n=0}^{N-1}(2n+\mu+1)\int_0^1
  dx x^{\mu+m}P_n^{(\mu,0)}(1-2x)P_n^{(\mu,0)}(1-2x)
\end{equation}
After the change of variables $x=(1-t)/2$ and the definition of
Jacobi polynomials as:
\begin{equation}\label{DefinitionJacobiPolynomials}
  P_n^{(\mu,0)}(t)=\frac{1}{n!}\sum_{k=0}^n\frac{(-n)_k (\mu+n+1)_k
  (\mu+k+1)_{n-k}}{k!}\left(\frac{1-t}{2}\right)^k
\end{equation}
(where $(a)_k=\Gamma(a+k)/\Gamma(a)$ is a Pochhammer symbol), we
obtain:
\begin{align}\label{moments}
 \nonumber\langle\lambda_1^m\rangle &=\frac{1}{2^{\mu+m+1}}\sum_{n=0}^{N-1}\frac{2n+\mu+1}{n!}
\sum_{k=0}^n\frac{(-n)_k (\mu+n+1)_k
  (\mu+k+1)_{n-k}}{2^k~k!}\times\\
  &\times\int_{-1}^1
  dt~(1-t)^{\mu+m+k}P_n^{(\mu,0)}(t)
\end{align}
The integral above can be computed for $m>-\mu-1$ (\cite{GR},
formula 7.391.2) in terms of a hypergeometric function $_3
F_2(-n,\mu+n+1,\mu+m+k+1;\mu+1,\mu+m+k+2;1)$. Since the first
argument is a negative integer, the series gets truncated to give
eventually\footnote{We are grateful to Marcel Novaes for
suggesting significant simplifications in \eqref{eventually}.}:
\begin{equation}\label{eventually}
 \langle\lambda_1^m\rangle
 =\sum_{n=0}^{N-1}(2n+\mu+1)\sum_{k,\ell=0}^n\frac{g(k)g(\ell)}{\mu+m+k+\ell+1}
\end{equation}
where:
\begin{equation}\label{gkappa}
  g(\kappa)=(-1)^\kappa\binom{n}{\kappa}\binom{n+\mu+\kappa}{\mu+\kappa}
  \end{equation}

 Despite lacking the aesthetic appeal of subcases already
 considered in the literature \cite{novaes}\cite{savin2}, formula \eqref{eventually} is
 nevertheless valid for any $(N_1,N_2)$ and $m>-\mu-1$, and is fully non-perturbative.
 After implementing \eqref{eventually} in \textsc{Mathematica}$^\circledR$, one can check by direct inspection
 that:
\begin{enumerate}
  \item the formula \eqref{eventually} agrees with the approximate result in
  \cite{novaes} (valid for $N_1,N_2\gg 1$):
\begin{equation}\label{Novaes}
  \langle \lambda_1^m\rangle =
  (\mu+2N)\sum_{p=1}^m\binom{m-1}{p-1}(-1)^{p-1}c_{p-1}\left(\frac{N(\mu+N)}{(\mu+2N)^2}\right)^p
\end{equation}
  where $c_p=\frac{1}{p+1}\binom{2p}{p}$ (see Table \ref{comp}).
  \item The shot noise power $\langle P\rangle/P_0$, defined as $\langle \lambda_1\rangle-\langle
  \lambda_1^2\rangle$, can be extracted from
  \eqref{eventually}. Thanks to multiple cancellations, the result
  can be cast in the very simple form:
\begin{equation}\label{Shot Noise}
  \frac{\langle P\rangle}{P_0}=\frac{N^2 (\mu+N)^2}{(\mu+2N-1)(\mu+2N)(\mu+2N+1)}
\end{equation}
which agrees with the known exact result \cite{braun}\cite{savin}
(see also eq. \eqref{AverageShotNoiseKnown} below).
  \item The average conductance $\langle G\rangle/G_0=\langle \lambda_1\rangle$
  from \eqref{eventually} can be brought to the simple form:
\begin{equation}\label{Conductance}
  \frac{\langle G\rangle}{G_0}=\frac{N(\mu+N)}{\mu+2N}
\end{equation}
which agrees with the known result \cite{baranger}.
\end{enumerate}
\begin{table}[htb]
\begin{center}
\begin{tabular}{|c|c|c|c|c|}
  \hline
  $\mu$ &  $N$ & m & Exact \eqref{eventually} & Approximate \eqref{Novaes} \\
  \hline
  4 & 57 & 3 & 18.4240 & 18.4248 \\
  4 & 87 & 7 & 18.637 & 18.638\\
  12 & 47 & 19 & 6.7672 & 6.77002\\
  15 & 57 & 29 & 6.67909 & 6.68199\\
  25 & 75 & 59 & 6.34394 & 6.34704\\ \hline
\end{tabular}
\caption{Comparison between the moments $\langle
\lambda_1^m\rangle$ computed by Novaes \cite{novaes} and our exact
derivation \eqref{eventually}. Note that the normalization
$\int_0^1 dx~\rho_\beta(x;N_1,N_2)=N_1$ implies that the moments
are not constrained between $0$ and $1$.}\label{comp}
\end{center}
\end{table}
\section{A second derivation of the average density and higher-order correlation
functions}\label{sect3} In this section, we will derive an
alternative expression for the average density of transmission
eigenvalues and higher order correlation functions for
\emph{finite} $N_1$ and $N_2$ and $\beta=2,4$, starting from the
jpd \eqref{jpd}. Exploiting a variant of the Selberg integral
evaluated by Kaneko \cite{kaneko}, all correlation functions can
be expressed in terms of a hypergeometric function of a matrix
argument, instead of a determinant of a kernel as in \eqref{corr}
(for $\beta=2$).

Consider the joint probability density of transmission eigenvalues
\eqref{jpd}:
\begin{equation}\label{jpd2}
  P_\beta(\lambda_1,\ldots,\lambda_N)=\mathcal{N}_\beta^{-1}\prod_{j<k}|\lambda_j-\lambda_k|^\beta
  \prod_{i=1}^N
  \lambda_i^{\frac{\beta}{2}(|N_2-N_1|+1)-1},\qquad
  0\leq\lambda_j\leq 1,
\end{equation}
where $N=\min(N_1,N_2)$, $\beta=1,2,4$ and the normalization
constant is given by \eqref{NormConstant}.

The density of eigenvalues is given by the following multiple
integral:
\begin{equation}\label{Density of Eigenvalues}
  \rho_\beta(\lambda_1;N_1,N_2)=N\int_0^1\ldots\int_0^1~d\lambda_2\ldots
  d\lambda_N~P_\beta(\lambda_1,\ldots,\lambda_N),
\end{equation}
such that the normalization $\int_0^1 \rho_\beta(x;N_1,N_2)dx =N$
holds (where again $N=\min(N_1,N_2)$).

It turns out that the integral above can be evaluated without the
use of the Orthogonal Polynomial technique, which would lead to
the formula \eqref{spectralexact}, if one resorts to the following
extension of Selberg integral given by Kaneko \cite{kaneko}:
\begin{align}\label{kan}
\nonumber\int_{[0,1]^n} &\prod_{j=1}^n dx_j \prod_{j=1}^n
x_j^{\ell_1}(1-x_j)^{\ell_2} \prod_{\substack{1\leq i\leq n\\
1\leq k\leq
m}}(x_i-t_k)\prod_{j<k}|x_j-x_k|^\beta=\\
& C_1~_2
F_1^{(\beta/2)}\left(-n,\frac{2}{\beta}(\ell_1+\ell_2+m+1)+n-1;\frac{2}{\beta}(\ell_1+m);\{t_1,\ldots,t_m\}\right),
\end{align}
where $C_1$ is a known constant and $_2 F_1^{(\alpha)}$ is a
hypergeometric function of a matrix argument. Details about these
objects are provided in the appendix.

From \eqref{Density of Eigenvalues}, one has:
\begin{align}\label{Density2}
\nonumber\rho_\beta(\lambda_1;N_1,N_2)
&=\frac{N\lambda_1^{\frac{\beta}{2}\left(|N_2-N_1|+1\right)-1}}{\mathcal{N}_\beta}\int_{[0,1]^{N-1}}
d\lambda_2\ldots d\lambda_N
\prod_{j<k}|\lambda_j-\lambda_k|^\beta\\
  &\times\prod_{i=2}^N
  \lambda_i^{\frac{\beta}{2}(|N_2-N_1|+1)-1}.
\end{align}
Now, the Vandermonde coupling can be decomposed as:
\begin{equation}\label{Vandermonde}
  \prod_{j<k}|\lambda_j-\lambda_k|^\beta=\prod_{j<k,j=2}|\lambda_j-\lambda_k|^\beta\prod_{j=2}^N
  |\lambda_1-\lambda_j|^\beta,
\end{equation}
and, for $\beta=2,4$ the absolute value in all products is
immaterial. Hence:
\begin{align}\label{Density3}
\nonumber\rho_\beta(\lambda_1;N_1,N_2)
&=\frac{N\lambda_1^{\frac{\beta}{2}\left(|N_2-N_1|+1\right)-1}}{\mathcal{N}_\beta}\int_{[0,1]^{N-1}}
d\lambda_2\ldots d\lambda_N
\prod_{j<k,j=2}|\lambda_j-\lambda_k|^\beta\\
 &\times \prod_{i=2}^N
  \lambda_i^{\frac{\beta}{2}(|N_2-N_1|+1)-1}\prod_{j=2}^N
  (\lambda_j-\lambda_1)^\beta.
\end{align}
Comparing \eqref{Density3} and \eqref{kan}, after the following
identification:
\begin{displaymath}
\left\{ \begin{array}{ll}
n &= N-1\\
\ell_1 &= \frac{\beta}{2}(|N_2-N_1|+1)-1\\
\ell_2 &=0\\
t_k &=\lambda_1\qquad\forall k=1,\ldots,m\\
m &=\beta
\end{array} \right.
\end{displaymath}
one eventually obtains:
\begin{align}\label{Density4}
\nonumber\rho_\beta(\lambda_1 &;N_1,N_2)
=\frac{N~C_1~\lambda_1^{\frac{\beta}{2}\left(|N_2-N_1|+1\right)-1}}{\mathcal{N}_\beta}
\\
&\times~_2
F_1^{(\beta/2)}\left(1-N,|N_2-N_1|+N+1;|N_2-N_1|+3-2/\beta;\lambda_1\mathbf{1}_{\beta}\right),
\end{align}
where we have introduced a customary matrix notation in the last
argument of the hypergeometric function. Note that the result
\eqref{Density4} is still formally valid for any even integer
$\beta$.

We also observe that higher order correlation functions can be
easily written down, exploiting the very same eq. \eqref{kan}. For
example, the two-point function
$\rho_\beta^{(2)}(\lambda_1,\lambda_2;N_1,N_2)$ can be written
(ignoring prefactors) as:
\begin{align}\label{2-point function}
\nonumber \rho_\beta^{(2)}(\lambda_1,\lambda_2 &;N_1,N_2) \propto
(\lambda_1
\lambda_2)^{\frac{\beta}{2}\left(|N_2-N_1|+1\right)-1}|\lambda_2-\lambda_1|^\beta\\
\nonumber &\times\int_{[0,1]^{N-2}} d\lambda_3\ldots
d\lambda_N\prod_{i=3}\lambda_i^{\frac{\beta}{2}\left(|N_2-N_1|+1\right)-1}
\prod_{j<k,j=3}|\lambda_j-\lambda_k|^\beta\\
&\times\prod_{j=3}^N |\lambda_j-\lambda_1|^\beta~\prod_{j=3}^N
|\lambda_j-\lambda_2|^\beta
\end{align}
and the $(N-2)$-fold integral is again of the same form as
\eqref{kan} for the following values of parameters:
\begin{displaymath}
\left\{ \begin{array}{ll}
n &= N-2\\
\ell_1 &= \frac{\beta}{2}(|N_2-N_1|+1)-1\\
\ell_2 &=0\\
t_k &=\lambda_1\qquad \forall k=1,\ldots,\beta\\
t_k &=\lambda_2\qquad\forall k=\beta+1,\ldots,2\beta\\
m &=2\beta
\end{array} \right.
\end{displaymath}
Hence, this time the matrix argument of the hypergeometric
function is
$\mathbf{X}^{(2)}:=\mathrm{diag}\left(\underbrace{\lambda_1,\ldots,\lambda_1}_{\beta\mbox{
times}},\underbrace{\lambda_2,\ldots,\lambda_2}_{\beta\mbox{
times}}\right)$. Note that the 2-point correlation function
$\rho_\beta^{(2)}(\lambda_1,\lambda_2;N_1,N_2)$ is manifestly
symmetric in the two arguments as it should, due to the symmetry
of Jack polynomials (see appendix). It is worth mentioning that
higher order correlation functions can be written down easily
along the same lines.

Thanks to a very efficient \textsc{Matlab}$^\circledR$
implementation of this kind of hypergeometric functions by Koev
and Edelman \cite{koev}, the density itself, linear statistics
(one-dimensional integrals over the density) and $n$-th order
correlation functions can be numerically tackled very easily. In
particular, these results entirely avoid the use of (quaternion)
determinants and (skew-)orthogonal polynomials which would arise
from the canonical RMT treatment and can get computationally
demanding for high $N_{1,2}$ and $n$. Conversely, the
computational complexity of the algorithm in \cite{koev} is only
\emph{linear} in the size of the matrix argument $(\beta n)$.

\begin{figure}[htb]
\begin{center}
\includegraphics[bb =52 196 548 591,totalheight=0.45\textheight]{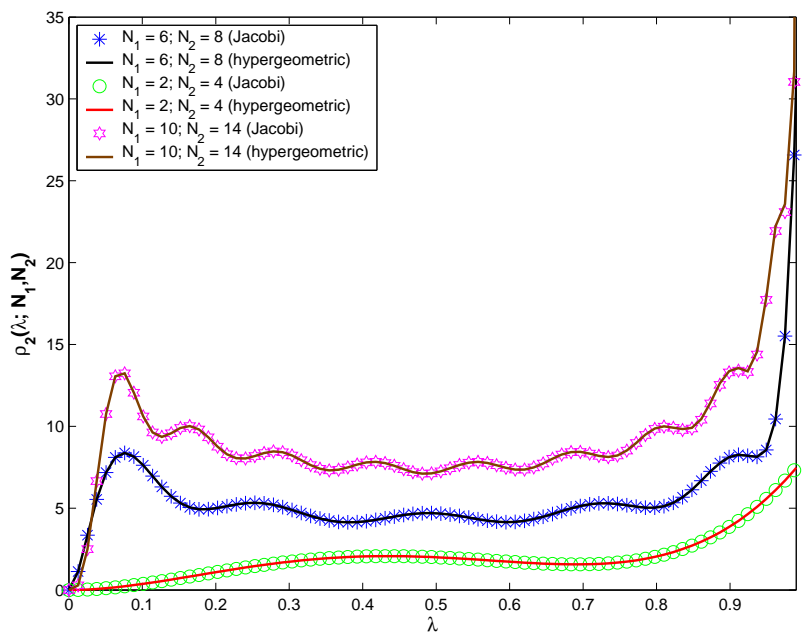}
\caption{Density of transmission eigenvalues for $\beta=2$ and
different values for the pair $(N_1,N_2)$. The plot symbols are
used for RMT formula \eqref{spectralexact}, whereas solid lines
represent the alternative formula \eqref{Density4}.
\label{DensityDodo}}
\end{center}
\end{figure}

In the following, we shall provide some plots of the spectral
density for different numbers of incoming and outgoing leads, and
$\beta=2$ (Fig. \ref{DensityDodo}). The agreement between the two
alternative formulas \eqref{spectralexact} and \eqref{Density4} is
excellent.

As a final check, we also numerically compute the prototype of
linear statistics, i.e. the (normalized) average shot noise power
$\langle P\rangle/P_0$ (see \eqref{AverageShotNoise}), defined as:
\begin{equation}\label{AverageShotNoisePower}
\langle P\rangle/P_0=\int_0^1 dx\rho_\beta(x;N_1,N_2)x(1-x)
\end{equation}
where $\rho_\beta(x;N_1,N_2)$ is taken from \eqref{Density4}. The
result has to agree with the analytical expression
\cite{braun}\cite{savin}:
\begin{equation}\label{AverageShotNoiseKnown}
  \frac{\langle P\rangle}{P_0}=\frac{N_1(N_1-1+2/\beta)N_2(N_2-1+2/\beta)}{(\tilde{N}-2+2/\beta)(\tilde{N}-1+2/\beta)(\tilde{N}-1+4/\beta)}
\end{equation}
where $\tilde{N}=N_1+N_2$. We compare in Table \ref{tab2} the
theoretical result \eqref{AverageShotNoiseKnown} with the
numerical integration of \eqref{AverageShotNoisePower}, obtained
in \textsc{MATLAB}$^\circledR$ with a standard integration
routine. The agreement we found is excellent, thus confirming the
correctness of \eqref{Density4}.

\begin{table}[htb]
\begin{center}
\begin{tabular}{|c|c|c|c|c|}
  \hline
  $N_1$ &  $N_2$ & $\beta$ & Theory & Numerical \\
  \hline
  4 & 7 & 2 & $0.5939393$ & $0.5939393$ \\
  8 & 11 & 2 & $1.1321637$ & $1.1321639$ \\
  3 & 9 & 2 & $0.4248251$ & $0.4248251$\\
  4 & 7 & 4 & $0.5805422$ & $0.5805424$ \\
  3 & 5 & 4 & $0.4326923$ & $ 0.4326923$ \\ \hline
\end{tabular}
\caption{Comparison between the theoretical expression for the
average shot-noise power \eqref{AverageShotNoiseKnown} and the
numerical integration of \eqref{AverageShotNoisePower}, for
different values of $N_1$, $N_2$ and $\beta$.}\label{tab2}
\end{center}
\end{table}
\section{Conclusions}
We have clarified that the average density of transmission
eigenvalues for chaotic cavities is analytically known from the
Jacobi ensemble of random matrices, as well as all higher-order
correlation functions. The known formula for the average density
coincides with the one derived with a mapping to a non-linear
sigma model by Ara\'{u}jo and Mac\^{e}do, but is rigorously valid
for any number of open channels. With an elementary integration
over this density, we derived a general formula for the moments
$\langle \lambda_1^m\rangle$, which is non-perturbative and valid
for arbitrary large $m$ and $\beta=2$. Also, a second derivation
is offered for the spectral density and higher order correlation
functions, which does not make use of orthogonal polynomials or
determinantal structures. Thanks to a recent algorithmic progress,
this result, exploiting a hypergeometric function of a matrix
argument, may be numerically easier to implement than high-order
(quaternion) determinants. All the results are consistent with
numerical checks and known formulas in the literature.
\section*{Acknowledgments}
PV has been supported by a Marie Curie Early Stage Training
Fellowship (NET-ACE project). We are indebted with Gernot Akemann
and Dmitry Savin for clarifying discussions and useful
suggestions. We are grateful to Piet Brouwer, Marcel Novaes and
Victor A. Gopar for pointing out relevant references. We also
thank Satya N. Majumdar and Oriol Bohigas for collaboration on
related topics and many helpful advices.
\appendix
\section{Hypergeometric function of a matrix argument}
Following Kaneko \cite{kaneko}, we first report the definition of
the constant $C_1$ appearing in \eqref{kan}:
\begin{equation}\label{C1}
  C_1:=S_{n,0}(\ell_1+m,\ell_2,\beta),
\end{equation}
where:
\begin{equation}\label{SelbKaneko}
   S_{n,0}(y_1,y_2,z):=\prod_{i=1}^n \frac{\Gamma\left(i\frac{z}{2}+1\right)\Gamma\left(y_1+1+(i-1)\frac{z}{2}\right)\Gamma\left(y_2+1+(i-1)\frac{z}{2}\right)}
  {\Gamma\left(\frac{z}{2}+1\right)\Gamma\left(y_1+y_2+2+(n+i-2)\frac{z}{2}\right)}.
\end{equation}

The hypergeometric function of a matrix argument \cite{murj} takes
a symmetric matrix $(m\times m)$ $\mathbf{X}$ as input and
provides a real number as output. It is defined as a series of
Jack functions of parameter $\beta$, which generalize the Schur
function, the zonal polynomial and the quaternion zonal polynomial
to which they reduce for $\beta=1,2,4$ respectively. Given a
partition $\kappa$ of an integer $k$, i.e. a set of integers
$\kappa_1\geq\kappa_2\geq\ldots\geq 0$ such that
$|\kappa|=\kappa_1+\kappa_2+\ldots=k$, and a matrix $\mathbf{X}$,
the Jack function $C_\kappa^{(\beta)}(\mathbf{X})$ is a symmetric
and homogeneous polynomial of degree $|\kappa|$ in the eigenvalues
$x_1,\ldots,x_m$ of $\mathbf{X}$.

The hypergeometric function is defined as:
\begin{equation}\label{hypergeometric}
  _p
  F_q^{(\beta)}(a_1,\ldots,a_p;b_1,\ldots,b_q;\mathbf{X}):=\sum_{k=0}^\infty
  \sum_{\kappa\vdash k}\frac{(a_1)_\kappa^{(\beta)}\ldots (a_p)_\kappa^{(\beta)}}{k!(b_1)_\kappa^{(\beta)}\ldots (b_q)_\kappa^{(\beta)}}
  C_\kappa^{(\beta)}(\mathbf{X}),
\end{equation}
where the symbol $\kappa\vdash k$ means that $\kappa$ is a
partition of $k$ and
$(a)_\kappa^{(\beta)}=\prod_{(i,j)\in\kappa}\left(a-\frac{i-1}{\beta}+j-1\right)$
is a generalized Pochhammer symbol.

The series \eqref{hypergeometric} converges for any $\mathbf{X}$
if $p\leq q$; it converges if $\max_i |x_i|<1$ and $p=q+1$; and
diverges if $p>q+1$, unless it terminates
\cite{koev}\cite{murj}\cite{forrester}.

\section{The $\mu\to 0$ limit of the spectral density.}
In the case $N_1=N_2=N$ and $\beta=2$, the average spectral
density was computed exactly in \cite{baranger} as:
\begin{equation}\label{SpectralDensityBaranger}
  \rho_2(\lambda;N_1=N_2=N)=\frac{N^2}{4\lambda(1-\lambda)}\{P_N^2(\alpha)-2\alpha P_N(\alpha)P_{N-1}(\alpha)
  +P_{N-1}^2(\alpha)\}
\end{equation}
where $\alpha=2\lambda-1$ and $P_N(x)$ is a Legendre polynomial.

This case corresponds to $\mu\to 0$ in eq. \eqref{spectralexact}.
In this appendix, we show explicitly how to get from
\eqref{spectralexact} to \eqref{SpectralDensityBaranger}.

First, we remark that the identity between Jacobi and Legendre
polynomials $P_n^{(0,0)}(x)=P_n(x)$ holds [formula 22.5.35 in
\cite{Abramo}]. Hence, in the case $\mu\to 0$ we have from
\eqref{spectralexact}:
\begin{equation}\label{spectrallimiting}
  \rho_2(\lambda;N_1=N_2=N)=\sum_{n=0}^{N-1}(2n+1)\{P_n(1-2\lambda)\}^2
\end{equation}
Next, we use the Christoffel-Darboux formula for Legendre
polynomials [see formula 22.12.1 in \cite{Abramo}] at equal
arguments $x=y=1-2\lambda$:
\begin{equation}\label{ChristoffelDarboux}
  \sum_{n=0}^{N-1}(2n+1)P_n^2(y)=N[P_N^\prime(y)P_{N-1}(y)-P_{N-1}^\prime (y)P_N(y)]
\end{equation}
Then, we exploit the differential relation [22.8.5 in
\cite{Abramo}]:
\begin{equation}\label{diff rel}
  (1-y^2)P_n^\prime (y)=-n y P_n(y)+n P_{n-1}(y)
\end{equation}
to get:
\begin{align}\label{spectrallimiting2}
  \nonumber\rho_2(\lambda;N_1=N_2=N) &=\frac{N}{1-y^2}[N P_{N-1}^2(y)-y P_N(y) P_{N-1}(y)-(N-1)\times\\
  &\times P_N(y) P_{N-2}(y)]\Big |_{y=1-2\lambda}
\end{align}
Thanks to the recurrence relation [22.7.10 in \cite{Abramo}], we
obtain the following identity for $P_{N-2}(y)$:
\begin{equation}\label{Pnminus2}
  P_{N-2}(y)=\frac{1}{N-1}[(2N-1)y P_{N-1}(y)-N P_N(y)]
\end{equation}
which is then substituted into \eqref{spectrallimiting2}.
Eventually, given that $y=1-2\lambda$ and the Legendre polynomials
have the same parity of their index, we obtain:
\begin{align}\label{eventually2}
  \nonumber\rho_2(\lambda;N_1=N_2=N) &=\frac{N^2}{4\lambda(1-\lambda)}[P_{N}^2(2\lambda-1)-2(2\lambda-1) P_N(2\lambda-1)\times\\
  &\times P_{N-1}(2\lambda-1)+P_{N-1}^2(2\lambda-1)]
\end{align}
in complete agreement with \eqref{SpectralDensityBaranger}.

\end{document}